\documentclass[preprint,12pt]{elsarticle}

\usepackage{amssymb}

 \usepackage{dcolumn}% Align table columns on decimal point
 \usepackage{bm}
 \newcommand     {\beq}[1]         { \begin{equation} #1 \end{equation} }
 \newcommand     {\beqa}[1]        { \begin{eqnarray} #1 \end{eqnarray} }

\usepackage{xcolor}

\journal{Chaos, Solitons, and Fractals}

\begin{document}

\begin{frontmatter}

\title{Transition from localized to mean field behaviour of cascading failures in the fiber bundle model on complex networks}

\author[unideb]{Attia Batool}
\author[unideb,atomki]{Gerg\H o P\'al}
\author[unideb]{Zsuzsa Danku}
% \author[unideb]{Vikt\'oria K\'ad\'ar}
% \author[kobe]{Naoki Yoshioka}
% \author[kobe]{Nobuyasu Ito}
\author[unideb,atomki]{Ferenc Kun}
\affiliation[unideb]{organization={University of Debrecen, Faculty of Science and Technology, 
Doctoral School of Physics, Department of Theoretical Physics},
            addressline={P.O.\ Box 400}, 
            city={Debrecen},
            postcode={H-4002}, 
            country={Hungary}}
\affiliation[atomki]{organization={Institute of Nuclear Research (Atomki)},
            addressline={Poroszlay \'ut 6/c}, 
            city={Debrecen},
            postcode={H-4026}, 
            country={Hungary}}
% \affiliation[kobe]{organization={RIKEN Center for Computational Science},
%             addressline={7-1-26 Minatojima-minami-machi, Chuo-ku, Hyogo}, 
%             city={Kobe},
%             postcode={650-0047}, 
%             country={Japan}}

\begin{abstract}
We study the failure process of fiber bundles on complex networks focusing on the effect of 
the degree of disorder of fibers' strength on the transition from localized to mean field 
behaviour. 
Starting from a regular square lattice we apply the Watts-Strogatz rewiring technique to
introduce long range random connections in the load transmission network and analyze how the ultimate strength 
of the bundle and the statistics of the size of failure cascades change 
when the rewiring probability is gradually increased. 
Our calculations revealed that the degree of strength disorder of nodes of the network has a substantial effect
on the localized to mean field transition. 
In particular, we show that the transition sets on at a finite value of the rewiring
probability, which shifts to higher values as the degree of disorder is reduced. 
The transition 
is limited to a well defined range of disorder, so that there exists a threshold disorder of nodes' strength 
below which the randomization of the network structure does not provide any improvement neither of the 
overall load bearing capacity nor of the cascade tolerance of the system. At low strength disorder 
the fully random network is the most stable one, while at high disorder 
best cascade tolerance is obtained at a lower structural randomness.
Based on the interplay of the 
network structure and strength disorder we construct an analytical argument which provides a reasonable 
description of the numerical findings.
\end{abstract}

\begin{graphicalabstract}
\end{graphicalabstract}

\begin{highlights}
\item A transition from localized to mean field behaviour of the failure dynamics of fiber bundles
      is revealed as the structure of the network of load transmitting connections is varied.
\item The transition is limited to a well defined range of disorder of the failure strength of nodes (fibers) of
      the network.
\item A threshold disorder is determined below which the randomization of the network structure does not 
      provide any improvement neither of the load bearing capacity nor of the cascade tolerance of the system.
\item An optimal network structure is identified with the highest stability against cascading failure.
\end{highlights}

\begin{keyword}
cascading failure \sep fiber bundle model \sep Watts-Strogatz rewiring \sep disorder of failure strength



\end{keyword}

\end{frontmatter}

\section{Introduction}

Load redistribution following local damage often drives the cascading failure of connected 
elements in complex systems \cite{boccaletti_complex_2006,chaos_blackout_review_2007}. 
From crackling bursts accompanying materials breakdown, 
through the failure avalanches of transportation and communication networks, to the cascading 
blackouts of electric transmission grids, the statistical features 
of failure cascades, and the overall performance of the damaged 
system strongly depend on the structure of the underlying network of load transmitting connections 
between elements of the system \cite{dou_robustness_2010,moreno_cascade_2013,load_sharing_complex_2014,profile_network_2017, neural_cascade_2020,chaos_review_2021,cascade_csf_2016,frasca_csf_2021}. 
The interplay of the cascading dynamics and of the network topology has recently been studied 
using discrete models on various types of complex networks. In these approaches 
either the nodes \cite{load_sharing_complex_2014,chaos_review_2021} or the links 
\cite{nuno_hans_pre_2014} of the network are assumed to undergo 
a degradation process accompanied by a mechanism of load rearrangement on the intact elements 
which can give rise to cascades of failure events. Among these approaches the so-called 
fiber bundle model (FBM), widely used to study materials breakdown phenomena \cite{hidalgo_avalanche_2009,hansen2015fiber,bikas_review_frontiersrt_2020}, has proven very useful
since it grasps the essential mechanisms of the intermittent failure spreading 
yet being simple enough to offer 
analytic solutions in certain limiting cases \cite{Moreno_2002,moreno_fracture_2000,hidalgo_avalanche_2009}.

In the basic setup, an FBM is composed of a set of parallel fibers organized on a regular lattice \cite{hansen2015fiber,kun_extensions_2006}. 
Under a slowly increasing external load the fibers fail irreversibly when the local load on them 
exceeds their strength value, which is assumed to have a certain degree of randomness. 
Under the constraint of load conservation, the load dropped by the failed fiber gets redistributed 
over the remaining intact ones. Recently, two limiting cases of load sharing have been subject to intensive investigations both with a high practical relevance: in case of equal load sharing (ELS) all intact fibers receive the same fraction of load irrespective 
of their distance from the failed one, while for localized load sharing (LLS) only the intact 
nearest neighbors share equally the load of the broken element \cite{kloster_burst_1997,hidalgo_fracture_2002,moreno_fracture_2000,hidalgo_avalanche_2009,danku_PhysRevLett.111.084302}. In both cases the load increments can cause 
further breakings so that a single broken fiber may trigger an entire cascade of failure events. 
Due to the generality of this failure spreading mechanism, fibers of the model can easily be replaced by 
roads carrying traffic \cite{bikas_physica_2006,fbm_traffic_ijmpc_2008}, flow channels 
\cite{talbot_traffic_pre_2015}, or electric power stations \cite{power_fiber_2015,chaos_blackout_review_2007, chaos_powergrid_2016, biswas_epl_2019,frasca_csf_2021} 
on a high voltage transmission grid, making FBMs a basic modelling framework for cascading failure 
with widespread applications on complex networks \cite{bikas_review_frontiersrt_2020}.

During the past decades it has been shown that for a broad class of the distributions of 
fibers' strength, FBMs exhibit
universal behaviour with two distinct universality classes according to the range of load redistribution:
for long range load sharing (ELS class) the size of failure cascades proved to be power law distributed 
with a universal exponent $5/2$ \cite{kloster_burst_1997,moreno_fracture_2000,hidalgo_avalanche_2009} 
and the bundle has a 
finite asymptotic strength in the limit of large system sizes \cite{smith_asymptotic_1982,mccartney_statistical_1983,hansen2015fiber}. 
For equal load sharing conditions the fibers always keep the same load, no stress concentration can arise,
hence, ELS realizes the mean field limit of FBMs.
Under short range load sharing (LLS class) the distribution of cascade sizes is a significantly steeper (non universal) power law or exponential, and additionally, for large system sizes the ultimate strength 
of the bundle tends to zero \cite{hemmer_distribution_1992,hansen_burst_1994,raischel_local_2006,PhysRevE.87.042816}. 
Recently, LLS FBMS have been analyzed on complex networks where fibers were assigned to the nodes
and localized load sharing was realized along the links of the network \cite{d.-h.kim_universality_2005}.
Based on the statistics of cascade sizes and on the ultimate strength of the system, it was demonstrated 
for scale-free, Erd\H os-R\'enyi (ER) and Watts-Strogatz (WS) rewired networks that LLS FBMs on 
complex networks 
fall in the ELS universality class \cite{d.-h.kim_universality_2005}. Later on it was shown on a ring graph 
with two nearest neighbor links that adding a single random load transmitting connection to each fiber, 
the localized load sharing FBM exhibits ELS behaviour in terms of the size distribution of failure 
cascades and global strength \cite{divakaran_fibers_2007}. However, all these studies of FBMs on 
complex networks were limited to a high disorder of node strength and considered only networks with a high degree of randomness
in their structure even in the case when all fibers had the same degree.

Here we present a detailed numerical and analytical study of the transition of the failure process of 
the fiber bundle model from the LLS to the ELS universality class when an initially 
regular lattice of load transmitting connections is gradually randomized. Starting from a square 
lattice we apply the Watts-Strogats rewiring technique \cite{watts_strogatz_nature_1998,watts_networks_1999} to
introduce long range random connections and study how the critical load and strain of the bundle, furthermore,
the statistics of the size of failure cascades change when the rewiring probability is gradually increased
at different degrees of disorder of the strength of nodes (fibers). 
Our calculations revealed that the degree of strength disorder of nodes of the network has a substantial effect
on the transition. In particular, we show that the LLS-ELS transition sets on at a finite value of the rewiring
probability, which shifts to higher values as the degree of disorder is reduced. The transition 
is limited to a well defined range of disorder, i.e.\ there exists a threshold disorder of nodes' strength 
below which the randomization of the network structure does not provide any improvement neither of the 
overall load bearing capacity nor of the cascade tolerance of the system. Based on the interplay of the 
network structure and strength disorder we construct an analytical argument which provides a reasonable 
description of the numerical findings.

\section{Fiber bundle model on a rewired square lattice}
To study cascading failures we consider a bundle of parallel fibers which are assigned to the nodes 
of a complex network. The bundle is subject to a slowly increasing external mechanical load parallel 
to the fibers'
direction. To connect the model to the mechanics of materials, we assume that the fibers are linearly 
elastic up to a threshold load 
$\sigma_{th}$ where they break irreversibly. The Young modulus $E$ of fibers has a fixed value $E=1$,
however, their local strength $\sigma_{th}$ is a random variable sampled from a probability
density function $p(\sigma_{th})$. When a fiber fails its load has to be overtaken by the remaining 
intact fibers. We assume localized load sharing (LLS), i.e.\ load is redistributed along the links of 
the underlying load transmission networks. In the following details of the model construction are presented:
\begin{figure}
\begin{center}
\includegraphics[bbllx=10,bblly=470,bburx=270,bbury=600,scale=1.2]{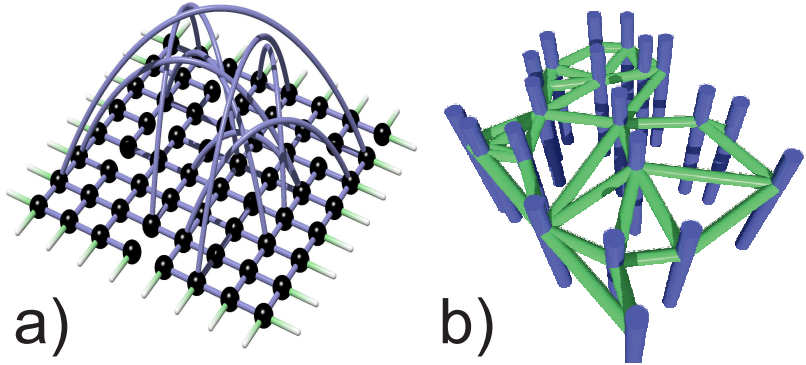}
  \caption{Demonstration of the model construction. $(a)$ The network of load transmitting connections 
  is obtained by rewiring a two-dimensional regular square lattice with periodic boundary connections
  in both directions. The rewiring introduces long range randomized connections, which broadens the degree
  distribution of the network while keeping the average degree fixed. $(b)$ Fibers of the bundle are assigned 
  to the nodes of the network oriented perpendicular to the plane of the original lattice. 
   \label{fig:model}}
\end{center}
\end{figure}

\begin{figure}
\begin{center}
\includegraphics[bbllx=0,bblly=0,bburx=350,bbury=305,scale=0.6]{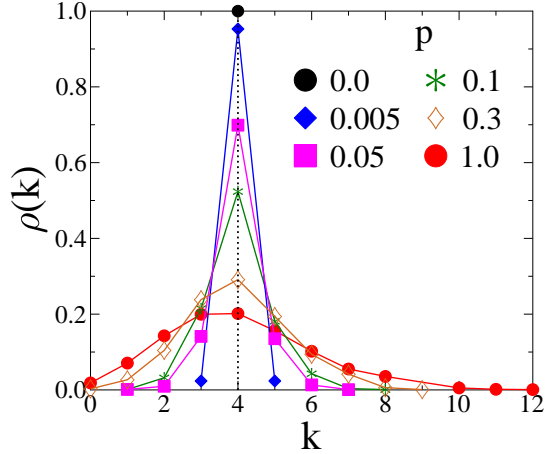}
  \caption{Degree distribution $\rho(k)$ of the network of fibers' connections at several rewiring probabilities $p$.
  As $p$ increases the distribution $\rho(k)$ gets broader, however, the value of the average number of connections is preserved $\left<k\right>=4$.
   \label{fig:degree}}
\end{center}
\end{figure}

To generate the network of connections along which load is redistributed over fibers, we start from a 
regular square lattice of side length $l$ with $N=l^2$ fibers, and apply the Watts-Strogatz 
rewiring technique to randomize the connections \cite{watts_strogatz_nature_1998,watts_networks_1999}.
The fibers are assigned to the nodes oriented perpendicular to the plane of the lattice. Figure \ref{fig:model}
provides an overview of the model construction.
On the square lattice with periodic boundary condition in both directions, all fibers (nodes) are connected 
to their four nearest neighbors, hence, initially the degree distribution $\rho(k)$ of fibers has the simple form
\begin{eqnarray}
\rho(k) = \left\{
\begin{array}{lcl}
1  & \mbox{for} & k=4, \\ [2mm]
0  &    \mbox{otherwise.} &
\end{array}
\right.
\end{eqnarray}
As to the next, each of the $L=2N$ initially existing connections is rewired with a probability $p$ 
which spans the interval $0\leq p \leq 1$. For both ends of a rewired link a new fiber is selected randomly
in the bundle with the constraint that neither multiple links nor loops are allowed between fibers 
(see Fig.\ \ref{fig:model} for illustration). As a consequence,
long range randomized connections are introduced in the bundle and the degree distributions $\rho(k)$ 
broadens while the average degree of nodes $\left<k\right>$, i.e.\ the average number of interacting 
partners of fibers, remains the same $\left<k\right>=4$. The degree distribution $\rho(k)$ of the network
is presented in Fig.\ \ref{fig:degree} for several values of the rewiring probability $p$. 
For large values of $p$ in the vicinity of 1, isolated fibers and small clusters of a few fibers 
may occur due to rewiring.
In order to exclude their effect, after the rewiring process we identify all clusters of nodes of the 
bundle and keep the largest one for further calculations.

\begin{figure}
\begin{center}
\includegraphics[bbllx=0,bblly=0,bburx=350,bbury=305,scale=0.6]{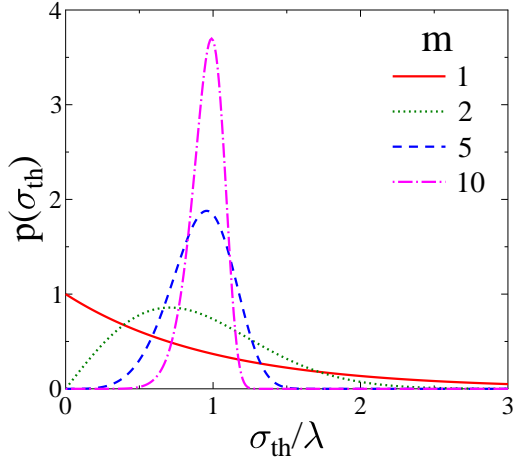}
  \caption{Weibull distribution of failure thresholds $p(\sigma_{th})$ of fibers 
  (nodes) of the network at different values of the exponent $m$. As the exponent $m$ 
  increases the distribution gets narrower.
   \label{fig:weibull}}
\end{center}
\end{figure}

The load bearing capacity of nodes, i.e.\ the threshold load $\sigma_{th}$ where fibers fail,
is a random variable which is sampled from a Weibull distribution 
\begin{equation}
 p(\sigma_{th})=m \frac{\sigma_{th}^{m-1}}{\lambda^m} e^{-\left(\sigma_{th}/\lambda\right)^m}
 \label{eq:expon_threshold}
\end{equation}
over the interval $0\leq\sigma_{th}<+\infty$.
Here the parameter $\lambda$ sets the scale of strength values, while $m$ controlls the shape 
of the distribution. 
The choice of the distribution Eq.\ (\ref{eq:expon_threshold}) has two motivations: 
$(i)$ the failure behaviour of FBMs with such a fast
decaying strength distribution, shows a high degree of
robustness which has been well understood both in the equal load sharing and localized
load sharing limits on regular square lattices \cite{andersen_tricritical_1997,hansen2015fiber}. 
$(ii)$ Varying the Weibull
shape parameter $m$ in the range $m\geq 1$, the degree of strength disorder can be controlled in the sense
that increasing $m$ reduces the width of the distribution making the response of the bundle more brittle.
This feature of the distribution is illustrated in Fig.\ \ref{fig:weibull} for several $m$ values.
It is an important characteristics of our model that the strength $\sigma_{th}^i$ and the degree $k_i$ 
of fibers (nodes) ($i=1,\ldots, N$) are uncorrelated.

As the load gradually increases on the bundle, initially all fibers keep the 
same load, hence, the weakest fiber with the lowest breaking threshold breaks first. 
We assume that fibers have a nearest neighbor
interaction so that the load dropped by a broken fiber is equally shared by its intact nearest neighbors
on the underlying network. As a consequence, the updated load of the neighboring fibers may exceed their local 
breaking threshold resulting 
in additional breakings which are then followed again by load redistribution. As a results of subsequent 
breaking and load redistribution steps a single fiber failure can trigger an entire cascade of failures,
which stops when all the fibers receiving load in a load redistribution step, can sustain the elevated 
load. This so called localized load sharing has the consequence that fibers breaking in an avalanche form a 
connected cluster on the underlying network in such a way that on the intact fibers along the cluster 
perimeter a large amount of load can accumulate. 

The system has two sources of disorder, i.e.\ the stochastic strength of fibers and the 
randomness of the underlying network of connections, which are both quenched. The interplay 
of the two gives rise to an inhomogeneous stress field on the fibers, which evolves as the failure 
of the system proceeds. If fiber $i$ of load $\sigma_i$ fails, then its $n_i$ intact nearest 
neighbors all receive the load increment $\Delta \sigma_i=\sigma_i/n_i$, so that 
the load $\sigma_j$ of a neighboring fiber $j$ will have the updated value
\begin{equation}
\sigma_j \to \sigma_j + \Delta \sigma_i. 
\end{equation}
It follows that $n_i\leq k_i$, where $k_i$ is the initial degree of the node $i$.
During an avalanche the external load is kept constant
so that the failure spreading is solely driven by the redistribution of load through the 
transmission network.

After a cascade stops the external load is further increased to provoke the breaking of a single element:
the load $\sigma_i$ of each intact fiber is incremented by the same amount $\delta \sigma$ 
\begin{equation}
\sigma_i \to \sigma_i + \delta \sigma, 
\end{equation}
where $\delta \sigma$  is determined as the smallest difference between the load $\sigma_i$ and 
strength $\sigma_{th}^i$ of intact fibers $\delta \sigma=\min\limits_i(\sigma_{th}^i-\sigma_i)$.
Ultimate failure of the system occurs when a load increment triggers a catastrophic cascade breaking 
all the intact fibers.

Simulations of the failure dynamics were performed starting from a square lattice 
of size $l=400$ with $N=160.000$ fibers using periodic boundary condition in both directions,
which results in $L=320.000$ load transmitting connections. 
To controll the degree of strength disorder, the scale parameter of the Weibull distribution 
was fixed to $\lambda=1$, while the shape parameter 
$m$ was varied in the range $1\leq m \leq 22$.
For the rewiring probability $p$ we considered 30 different values in the interval $0\leq p \leq 1$.
At each parameter set averages were calculated over 2000 samples.

\section{Macroscopic response of fiber bundles on complex networks}
\begin{figure}
\begin{center}
\includegraphics[bbllx=20,bblly=15,bburx=380,bbury=330,scale=0.5]{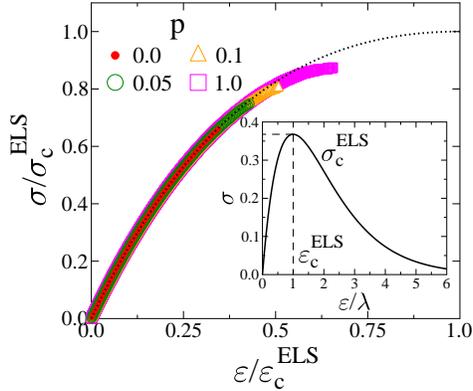}
  \caption{Constitutive curves $\sigma(\varepsilon)$ of the bundle at different rewiring probabilities $p$ 
  compared to the mean field solution of the model Eq.\ (\ref{eq:constit_example}) indicated by the dotted line for the Weibull parameter $m=1$.
  The two axis are rescaled by the mean field critical load $\sigma_c^{ELS}$ and strain $\varepsilon_c^{ELS}$.
  The inset demonstrates the entire $\sigma(\varepsilon)$ curve of Eq.\ (\ref{eq:constit_example}) including 
  the definition of the critical load $\sigma_c^{ELS}$ and strain $\varepsilon_c^{ELS}$.
   \label{fig:constit}}
\end{center}
\end{figure}
The macroscopic response of the bundle can be characterized by determining the relation 
$\sigma(\varepsilon)$ of its stress $\sigma$ and strain $\varepsilon$.  
In the limit of equal load sharing, where all fibers interact with each other and
keep the same load, this constitutive relation can be obtained analytically as
\begin{equation}
  \sigma = E\varepsilon\left[1-P(E\varepsilon)\right],
  \label{eq:constit_els}
\end{equation}
where $P(x)$ denotes the cumulative distribution of failure thresholds
\cite{kun_extensions_2006,hansen2015fiber}. Since at a given strain 
$\varepsilon$ all the fibers keep the same load $E\varepsilon$, the total load on the system is  
the product of the load of single fibers and of the fraction of intact fibers $1-P(E\varepsilon)$.
Substituting the Weibull distribution of thresholds Eq.\ (\ref{eq:expon_threshold}) we obtain 
the mean field constitutive equation of our model
\begin{equation}
 \sigma(\varepsilon) = \displaystyle{E\varepsilon e^{-\left(\frac{E \varepsilon}{\lambda}\right)^m}},
 \label{eq:constit_example}
\end{equation}
which is presented by the inset of Fig.\ \ref{fig:constit} for $m=1$. In a load 
controlled experiment the constitutive curve can only be realized up to the maximum, where immediate 
failure occurs in the form of a catastrophic cascade. 
Hence, the value $\sigma_c$ and the position $\varepsilon_c$ of the maximum define the critical load 
and strain of the bundle, respectively. 

In computer simulations of finite bundles of localized load sharing, the stress $\sigma$ and strain $\varepsilon$ of the system 
\begin{figure}
\begin{center}
\includegraphics[bbllx=20,bblly=15,bburx=380,bbury=330,scale=0.65]{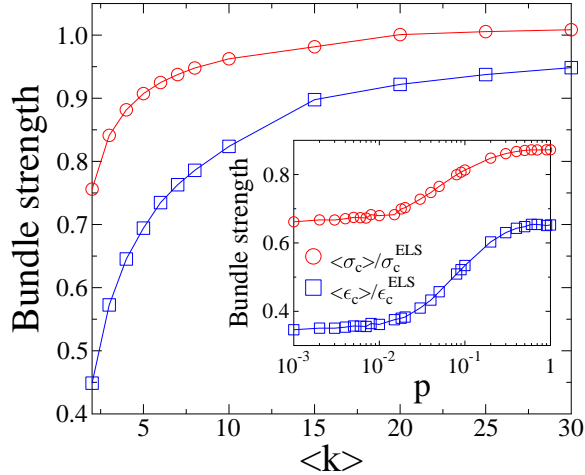}
  \caption{Inset: Average critical strain $\left<\varepsilon_c\right>$ and load $\left<\sigma_c\right>$ 
  of the network scaled with their mean field counterparts $\varepsilon_c^{ELS}$ and $\sigma_c^{ELS}$ 
  as function of the rewiring probability $p$ for the Weibull parameter $m=1$ of the failure thresholds.
  Main panel: The strength values $\left<\varepsilon_c\right>$ and  $\left<\sigma_c\right>$ obtained 
  for $p=1$ as function of the average number of neighbors in the range $2\leq \left<k\right> \leq 30$.
  The legend is provided in the inset for both figures.
   \label{fig:averstrength}}
\end{center}
\end{figure}
can be obtained on any network topology by summing up the load $\sigma_i$ of fibers (nodes) $F=\sum_{i=1}^N \sigma_i$ and dividing it by the bundle size $\sigma=F/N$ and by the total number of intact fibers $\varepsilon=F/N_{intact}$, respectively.
It has been shown by means of computer simulations
that for localized load sharing on regular lattices the constitutive curve of the bundle follows 
the mean field solution Eq.\ (\ref{eq:constit_els}) \cite{kun_damage_2000-1,hidalgo_fracture_2002}. 
However, the response becomes more brittle in the sense
that the $\sigma(\varepsilon)$ curve stops earlier at lower $\sigma_c$ and $\varepsilon_c$ 
closer to the initial linear regime.
Figure \ref{fig:constit} demonstrates that this behaviour remains valid for all the networks considered,
however, as the rewiring probability $p$ increases, the constitutive curves reach to higher $\sigma_c$ 
and $\varepsilon_c$ values.

To have a more transparent view on the effect of the network structure on the strength of the bundle,
we determined the average value of the critical load $\left<\sigma_c\right>$ 
and strain $\left<\varepsilon_c\right>$ as function of the rewiring probability $p$. 
It can be observed in the inset of Fig.\ \ref{fig:averstrength} for the Weibull parameter $m=1$ that 
for small values of $p$ the randomized contacts hardly have any effect on the strength of the bundle so that 
both the critical load $\left<\sigma_c\right>$ and strain $\left<\varepsilon_c\right>$ 
retain their original values characteristic for the square lattice at $p=0$. 
However, when the rewiring probability exceeds a threshold value $0.01\lesssim p$ the strength starts to increase 
and saturates to a limit value for completely randomized networks $p\to 1$. Note that in the figure the strength values are scaled with their mean field (equal load sharing) counterparts 
\beqa{
\varepsilon_c^{ELS}&=&\frac{\lambda}{E}\left(\frac{1}{m}\right)^{1/m}, \\ [2mm]
\sigma_c^{ELS}&=&\lambda \left(\frac{1}{m}\right)^{1/m}e^{-1/m},
\label{eq:els_crit}
}
obtained as the position and value of the maximum of the $\sigma(\varepsilon)$ curve of 
Eq.\ (\ref{eq:constit_example}).
This comparison shows that as random connections start to dominate the load transmission among fibers, 
the strength of the bundle approaches the equal load sharing limit but saturates at a lower value. 
The reason is that the average number 
of interacting partners of fibers is fixed to $\left<k\right>=4$ which still gives rise to a significant 
stress concentration on the network, and hence, reduces the fracture strength of the bundle 
compared to the mean field limit.
To support this argument we performed computer simulations on random graphs corresponding to the $p=1$ limit
of our system with the same number of nodes $N$ as the original square lattice varying the average number 
of neighbors $\left<k\right>$ in a broad range. Figure \ref{fig:averstrength} demonstrates that increasing 
the average number of interacting partners $\left<k\right>$ of fibers in a completely random network $p=1$ 
the strength values 
$\left<\sigma_c\right>$ and  $\left<\varepsilon_c\right>$ converge to their mean field counterparts 
as expected.

\section{Size distribution of failure cascades}
The  microscopic mechanism of the failure of the system is the cascading failure of nodes (fibers) 
triggered by single 
\begin{figure}
\begin{center}
\includegraphics[bbllx=5,bblly=40,bburx=370,bbury=310,scale=0.65]{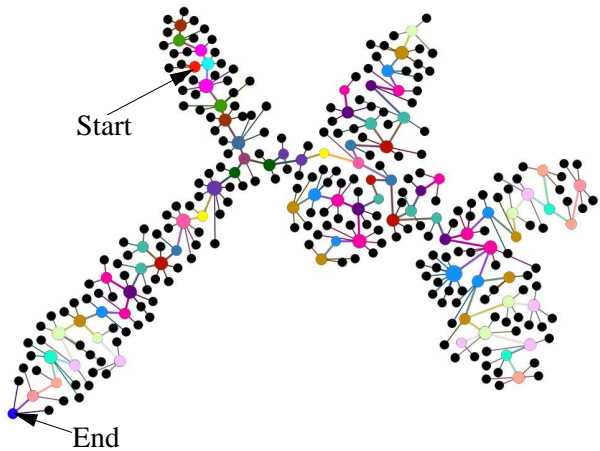}
  \caption{Spreading of a failure cascade on the network of load sharing connections at $p=0.5$ 
  with Weibull distributed failure thresholds at $m=1$. 
  The cascade starts 
  and ends with a single breaking indicated by the arrows. All fibers of the network which receive 
  load from broken fibers are indicated by black circles, and among them the ones which break as a consequence 
  of load sharing are highlighted by colors different from black. 
  For clarity, the circles representing broken fibers 
  have also a larger size. Fibers breaking in the same load redistribution step have the same color. The cascade in the example has the size $\Delta=87$, which was generated in 24 load redistribution steps. 
  For clarity, nodes of the network which do not participate in the cascade are not shown.
   \label{fig:avalspread}}
\end{center}
\end{figure}
breaking events as a consequence of external load increments. Our simulations revealed that 
the structure of the load transmission network plays an essential role in the growth of avalanches 
which in turn also determines the macroscopic behaviour of the
bundle. A cascade always starts from a single failing node and spreads over the transmission network, which is demonstrated in Fig.\ \ref{fig:avalspread} for a network at the rewiring probability $p=0.5$ with the Weibull shape 
parameter $m=1$ of strength values.
In the figure all fibers (nodes) which receive load from breaking fibers are indicated by black color, and among 
them those ones which suffer breaking are highlighted by colors different from black. Fibers breaking 
as a consequence of the same load redistribution step, are represented by the same color. 
Starting from the externally imposed fiber breaking 
in the top of the figure, one can easily follow the development of the cascade through the consecutive colors. Note that 
the cascade forms a connected cluster of broken fibers on the network, however, fibers breaking in the same
sub-cascade can be far from each other. The number of fibers breaking in the cascade defines the cascade size $\Delta$.
\begin{figure}
\begin{center}
\includegraphics[bbllx=30,bblly=30,bburx=380,bbury=330,scale=0.65]{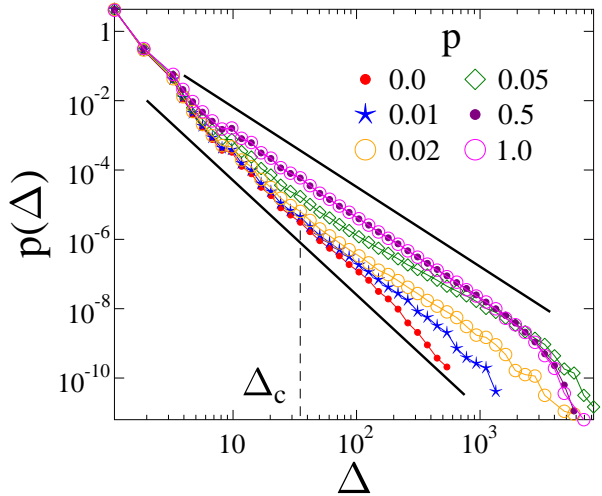}
  \caption{The size distribution of failure cascades $p(\Delta)$ for several values of the rewiring 
  probability $p$ at the Weibull parameter $m=1$ of the strength of fibers. 
  For low $p$ values a crossover occurs between two power law regimes of different exponents. 
  The two straight lines represent power laws of exponents $2.3$ and $3.4$, characteristic for $p=1$
  and $p=0$, respectively. The vertical dashed line highlights the crossover burst size $\Delta_c$ estimated for 
  $p=0.01$.
   \label{fig:avalsize}}
\end{center}
\end{figure}

To characterize the statistics of failure cascades we determined the probability distribution of their size $p(\Delta)$, which proved to have a strong dependence on the network topology of load transmitting connections. It can be observed in Fig.\ \ref{fig:avalsize} for the Weibull parameter $m=1$ 
that on the regular square lattice ($p=0$) where strong spatial localization of load occurs 
around failed regions, the distribution $p(\Delta)$ can be approximated as a power law 
\beq{
p(\Delta) \sim \Delta^{-\tau},
}
which is followed by a finite size cutoff. The value of the exponent is rather high $\tau=3.4\pm 0.1$ in
agreement with former studies of the fiber bundle model \cite{raischel_local_2006,kun_damage_2000-1,hidalgo_fracture_2002}. The rapidly decreasing distribution and the low cutoff burst size $\Delta_{max}$ clearly show that cascades are typically small compared to the system size $N$. Due to the strong localization large avalanches would lead 
to immediate collapse of the bundle on the regular lattice.
However, increasing the fraction of long range connections by increasing the rewiring probability $p$, 
the stress localization gets gradually reduced, hence, the system can tolerate larger and 
larger failure cascades without suffering catastrophic collapse. 
As a consequence, the cutoff burst size $\Delta_{max}$ increases and the distribution $p(\Delta)$ 
exhibits a crossover to a second power law regime 
with a lower exponent. The smaller value of $\tau$ shows the growing fraction of large size 
cascades in the failure dynamics of the system (see Fig.\ \ref{fig:avalsize}). 

\begin{figure}
\begin{center}
\includegraphics[bbllx=40,bblly=30,bburx=380,bbury=330,scale=0.65]{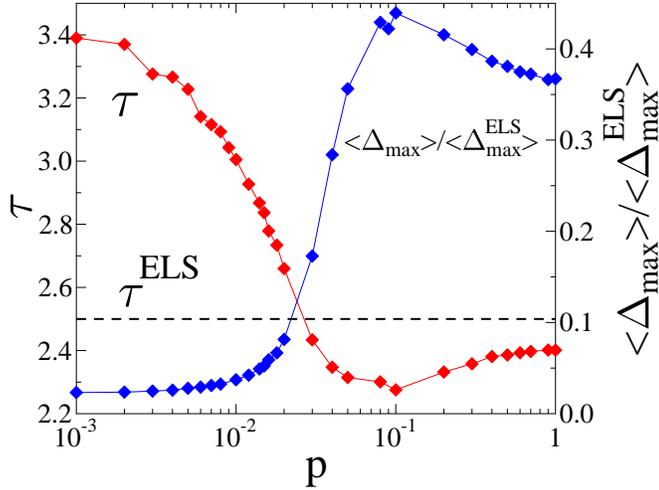}
  \caption{The power law exponent $\tau$ of the probability distribution of the cascade size
  $p(\Delta)$ as a function of the rewiring probability $p$ (left vertical axis). The average value of the largest cascade $\left<\Delta_{max}\right>$ scaled with the corresponding ELS result 
  $\left<\Delta_{max}^{ELS}\right>$ is presented on the right vertical axis as a function of $p$. The horizontal dashed line represents the mean field value of the cascade size  exponent $\tau^{ELS}=5/2$. 
  All the results were obtained at the Weibull parameter $m=1$.
   \label{fig:parameters}}
\end{center}
\end{figure}
Of course, the crossover burst size $\Delta_c$, which separates the two power law regimes,
depends on the rewiring probability, i.e.\ $\Delta_c$ gradually shifts to smaller values
with increasing $p$ in such a way that in the limit $p\to 1$ practically a single power law remains with a significantly lower exponent than that of the original square lattice at $p=0$. To characterize this evolution of $p(\Delta)$ we determined the average size of the largest avalanche 
$\left<\Delta_{max}\right>$ and the power law exponent $\tau$
of the regime of large avalanches as function of the rewiring probability $p$. 
It can be observed in Fig.\ \ref{fig:parameters} for the Weibull parameter $m=1$ 
that up to the rewiring probability $p_l\approx 0.01$
the cutoff cascade size $\left<\Delta_{max}\right>$ is nearly constant, although the exponent $\tau$ suffered some change. 
This behaviour implies that the small fraction of randomized contacts has a minor effect on the cascading 
failure dynamics in this parameter range $p\lesssim 0.01$. However, above this threshold probability
a rapid change of the cascade size distribution sets on indicated by the steep increase of $\left<\Delta_{max}\right>$ and decrease of the exponent $\tau$. For high values of the rewiring 
probability $p\to 1$, the exponent $\tau$ converges to a constant $\tau\approx 2.3$, 
which falls very close to the mean field burst size exponent of FBMs $\tau^{ELS}=5/2$ \cite{kloster_burst_1997,hidalgo_avalanche_2009}. The result
indicates that on sufficiently randomized load transmission networks the statistics of failure cascades
of the localized load sharing FBM becomes equivalent to the mean field universality class of the system
in agreement with the behaviour of the macroscopic strength of the bundle.
The result is consistent with Ref.\ \cite{d.-h.kim_universality_2005} where FBMs were analyzed 
on Watts-Strogatz networks in the range of high rewiring probabilities $p\geq 0.2$ recovering the mean 
field behaviour.

It is interesting to note that in Fig.\ \ref{fig:parameters} the average largest cascade size 
$\left<\Delta_{max}\right>$ has a maximum around the rewiring probability $p^*\approx 0.1$ which 
practically coincides with the position of the minimum value of the exponent $\tau$.
This behaviour indicates that there exists a network topology determined by $p^*$ where the network 
can tolerate the largest cascades with a considerable frequency. The reason is that increasing the rewiring probability the growing randomness of the network increases the perimeter of the failed clusters, 
hence, reducing the load concentration on it. This mechanism stabilizes the system in the sense that 
cascades can reach larger sizes without becoming instable destroying the system. However, at higher $p$ a counter 
effect occurs that low degree fibers appear on the network with a growing fraction, which increases the load concentration
in their vicinity and makes the system more vulnerable to cascades. 
The value $p^*$ provides the optimum for the cascade tolerance of the 
system.
\begin{figure}
\begin{center}
\includegraphics[bbllx=20,bblly=20,bburx=740,bbury=635,scale=0.50]{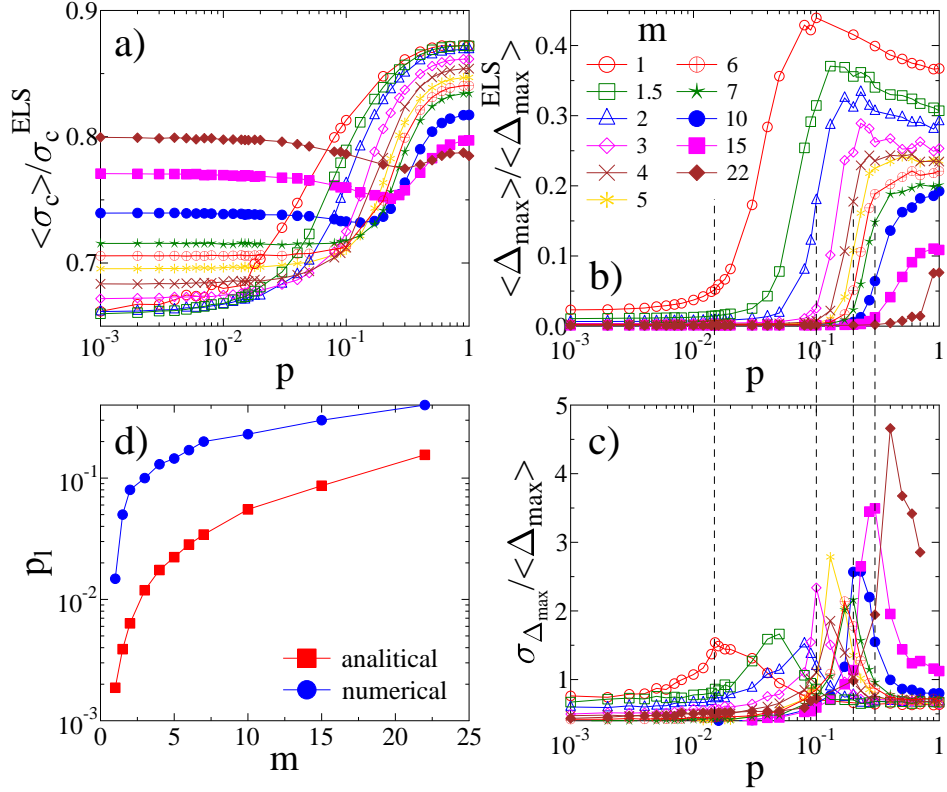}
  \caption{The average critical load $\left<\sigma_c\right>$ $(a)$ and the average size of the largest cascade $\left<\Delta_{max}\right>$ $(b)$ scaled with their mean field values $\sigma_c^{ELS}$ and $\left<\Delta_{max}^{ELS}\right>$ as function of the rewiring probability $p$ for several values of the Weibull parameter $m$.
  In $(b)$ the position of the local maximum of $\left<\Delta_{max}\right>$ determines the rewiring
  probability $p^*$ where the network tolerates the largest failure cascades. 
  $(c)$ The relative scatter of the size of the largest cascade, i.e.\ the standard deviation $\sigma_{\Delta_{max}}$ divided by the average $\left<\Delta_{max}\right>$ of the largest cascade size. The dashed lines serve to guide the eye indicating the position of the maximum of the curves 
  for the Weibull parameters $m=1, 3, 10, 15$. For $(a,b,c)$ the legend is provided in $(b)$. 
  $(d)$ The rewiring probability $p_l$ of the onset of the transition from LLS to ELS behaviour obtained as the position of the maximum of the curves in $(c)$ ({\it numerical}) and as the solution of Eq.\ (\ref{eq:analitic_pl}) ({\it analytical}).
   \label{fig:disorder_all}}
\end{center}
\end{figure}

% \begin{figure}
% \begin{center}
% \includegraphics[bbllx=20,bblly=15,bburx=380,bbury=330,scale=0.65]{aver_sigmac_all.eps}
% %\includegraphics[scale=0.8,origin=c]{mass_top2_feri.eps}
%   \caption{Average critical stress $\left<\sigma_c\right>$ as a function of the rewiring probability $p$ 
%   for several values of the Weibull parameter $m$.
%   The strength values are scaled with their mean field counterparts $\sigma_c^{MF}$.
%    \label{fig:aversigc_all}}
% \end{center}
% \end{figure}

\section{Effect of the degree of threshold disorder on the LLS-ELS transition}
We carried out a large amount of simulations of the failure process of the network of fibers at several values
of the Weibull exponent $m$ varying the degree of strength disorder in a broad range. 
These calculations revealed that the LLS to ELS transition has a high complexity as the network structure is
gradually changed where the degree of strength disorder of nodes plays a crucial role.
Figure \ref{fig:disorder_all}$(a)$ demonstrates for the average critical load 
$\left<\sigma_c\right>$ that as the strength disorder is reduced by increasing $m$, the onset of the transition, i.e.\ the rewiring probability $p_l$ where the first 
significant deviation occurs from the LLS result of the regular lattice, shifts to higher values. For instance,
for $m=7$ the transition starts at about $p_l\approx 0.1$, which is an order of magnitude higher 
than the corresponding value $p_l\approx 0.01$ obtained at $m=1$.
The transition is completed at a rewiring probability $p_u$ beyond which the bundle strength practically 
does not change. With decreasing strength disorder the value of $p_u$ also
increases and tends towards 1  in such a way that the transition regime shrinks. 
Note that the asymptotic strength $\left<\sigma_c\right>(p=1)$ decreases with increasing $m$ 
compared to its mean field counterpart $\sigma_c^{ELS}$, indicating that at lower threshold disorder randomization of 
the network structure provides less and less improvement of the overall load bearing capacity of the system. 
It is interesting to note that at the lowest disorders considered, starting from $m=10$, the $\left<\sigma_{c}\right>(p)$ curves proved 
to be non-monotonous, i.e.\ for $m=10, 15$ the onset of the increase of the ultimate strength of the system is preceded by a local minimum. Additionally, for $m=22$ the limit value of the strength attained at $p\to 1$ falls below the strength of the original square lattice. The result implies that when the strength of nodes is sampled from a sufficiently narrow interval the rewiring process gives rise to a reduction of the bundle strength at any rewiring probability.

% \begin{figure}
% \begin{center}
% \includegraphics[bbllx=30,bblly=30,bburx=380,bbury=330,scale=0.65]{maxburstsize.eps}
% %\includegraphics[scale=0.8,origin=c]{mass_top2_feri.eps}
%   \caption{The average size of the largest burst $\left<\Delta_{max}\right>$ scaled with the corresponding mean field value $\left<\Delta_{max}^{ELS}\right>$ as a function of the rewiring probability $p$ for several values of the Weibull shape parameter $m$. The position of the local maximum determines the rewiring
%   probability $p^*$ where the network tolerates the largest failure cascades.
%    \label{fig:maxburst_m}}
% \end{center}
% \end{figure}

The analysis of the statistics of cascade size revealed a similar effect of the strength disorder of fibers
on the LLS-ELS transition of the failure process: 
It can be observed in Fig.\ \ref{fig:disorder_all}$(b)$ for the average size of the largest cascade $\left<\Delta_{max}\right>$  that 
as the degree of threshold disorder gets reduced with increasing $m$ 
the value of $\left<\Delta_{max}\right>$ remains constant keeping its $p=0$ value 
for a broader and broader range of the rewiring probability $p$.
The estimated lower bounds $p_l$ of the transition regime are consistent in Figs.\  
\ref{fig:disorder_all}$(a)$ and \ref{fig:disorder_all}$(b)$ for the macroscopic and microscopic quantities 
showing that $p_l$ increases with decreasing threshold disorder. Note that the $\left<\Delta_{max}\right>(p)$
curves rise sharper than the bundle strength $\left<\sigma_c\right>(p)$ making the transition regime more
transparent. It can be expected that at the onset of the LLS-ELS transition the value of $\Delta_{max}$ 
has large fluctuations. To quantify this Fig.\ \ref{fig:disorder_all}$(c)$ presents the relative scatter of
$\Delta_{max}$, i.e.\ the ratio of its standard deviation $\sigma_{\Delta_{max}}$ and average 
$\left<\Delta_{max}\right>$. For each degree of disorder $m$, a sharp maximum can be observed whose position
provides a good measure of $p_l$. The vertical dashed lines highlight for a few Weibull exponents $m$ that 
indeed the maximum of the relative scatter of $\Delta_{max}$ well coincides with the onset of the sharp
rise of $\left<\Delta_{max}\right>(p)$ in Fig.\ \ref{fig:disorder_all}$(b)$.

It is important to note that as the strength disorder decreases the position $p^*$ of the maximum 
of $\left<\Delta_{max}\right>$ where the network tolerates the largest avalanches, shifts 
to higher values. Additionally, the maximum gradually decreases 
and eventually disappears around $m\approx 4$, where the $\left<\Delta_{max}(p)\right>$ curves become monotonous.
It follows that for threshold disorder in the range $m>4$ the fully random graph provides the highest 
tolerance of cascades. In agreement
with the behaviour of the ultimate strength of the bundle, at lower strength disorder of the nodes, 
the randomization of the network structure provides less and less improvement compared to the 
LLS limit of regular lattices. 

For each degree of disorder $m$ the size distribution of cascades $p(\Delta)$ goes over the same evolution
as for $m=1$ in Fig.\ \ref{fig:avalsize}: below the rewiring probability $p_l$ the cascade size distribution 
practically remains the same as on the original regular lattice at $p=0$. The second power law regime with a lower exponent emerges for networks with $p\geq p_l$ accompanied by the growth of the cutoff cascade size $\left<\Delta_{max}\right>$ and by the gradual decrease of the crossover cascade size $\Delta_c$. When $p$ exceeds 
$p_u$, the transition gets completed and a single power law remains of $p(\Delta)$.

\subsection{Rewiring probability of the onset of the transition}
In order to understand how the transition emerges from LLS to ELS with the rewiring probability $p$
at different degrees of disorder $m$, we construct an analytical argument based on the changing 
structure of the underlying 
load transmission network. On the original square lattice of fibers, localized 
load sharing dynamics leads to early failure of the entire bundle due to the strong stress concentration on 
the perimeter of failed clusters \cite{zapperi_analysis_1999,raischel_local_2006,kun_damage_2000-1}. 
In the last stable configuration of the bundle failed clusters are very small compared to the system size
so that the majority of fibers break in the last catastrophic cascade. Adding randomized long range 
connections leads to the reduction of the local stress concentration by increasing the perimeter of the 
growing failed cluster. As a consequence, the system can tolerate larger cascades and has a higher overall 
load bearing capacity.

At a given value of $p$ the average number of rewired connections 
can be estimated as $2Np$ since each of the initial $2N$ links is rewired with probability $p$.
When $p$ is low only a very small fraction of fibers is affected in the bundle by the rewiring 
either by having a removed nearest neighbor link or by getting a newly established long range contact.
So it is reasonable to assume that at low $p$ values the majority of spreading cascades have 
a high chance to avoid fibers with rewired links so that cascades remain small following 
the same statistics as on the original square lattice at $p=0$. Those cascades which involve 
fibers with rewired connections may grow to larger sizes resulting in a statistics different 
from the one of the small avalanches. This mechanism leads to the emergence of a crossover in the
distribution of the size of cascades presented in Fig.\ \ref{fig:avalsize}.

To estimate the crossover burst size $\Delta_c$, it is instructive to determine the probability 
that a randomly selected 
node of the network is affected by rewiring. The probability that none of the 4 nearest neighbor 
connections is rewired for a fiber is $(1-p)^4$, while the probability that it does not get connected 
to any new fiber can be estimated as $\exp{\left(-4p\right)}$ for large $N$ \cite{newman_structure_2003}. 
Hence, the probability $p_r$
that a node is affected by rewiring can be cast into the form
\beq{
p_r=1-(1-p)^4 e^{-4p}.
}
We assume that crossover occurs at a cascade size $\Delta_c$ above which cascades involve on average 
at least one fiber affected by the rewiring process. Hence, the relation 
\beq{
\Delta_c p_r(p) \approx 1
}
follows between $\Delta_c$ and the rewiring probability $p$, from which
we obtain
\beq{
\Delta_c \approx \frac{1}{p_r(p)}.
\label{eq:crossover}
}
Note that for $p\to 0$ the crossover avalanche size diverges $\Delta_c\to\infty$, while it tends to 1 for $p\to 1$.

It follows from the above arguments that at very low rewiring probabilities $p\ll 1$ the crossover 
cascade size is larger than the average largest avalanche $\left<\Delta_{max}\right>(p=0)$ 
on the original regular lattice at $p=0$. It has 
the consequence  that the dynamics and statistics of stable cascades is practically 
not affected by the rewiring in this 
$p$ range, no crossover occurs, so that the distributions of the cascade size remain practically 
the same as on the square lattice (see Fig.\ \ref{fig:avalsize}).
Crossover of the distributions emerges for those $p$ values where the condition
\beq{
\left<\Delta_{max}\right>(p=0, m)>\Delta_c(p)
}
holds. Note that $\left<\Delta_{max}\right>$ of the original lattice $p=0$ also depends on the 
strength disorder $m$. This first occurs at the lower bound $p_l$ of the transition regime
\beq{
\left<\Delta_{max}\right>(p=0,m) = \Delta_c(p_l)
\label{eq:analitic_pl}
}
from which $p_l$ can be obtained as a function of the degree of strength disorder $p_l=p_l(m)$.
In the regime $p>p_l(m)$ the crossover cascade size separating the two power law regimes of 
different exponents can be approximated by Eq.\ (\ref{eq:crossover}).
The smallest possible value of $\Delta_c$ we could identify in our numerical measurements 
is $\Delta_c\approx 1-3$, from which the upper bound of the crossover $p_u$ can be determined.
Further increasing $p$ above $p_u$ no qualitative 
change of the failure process occurs so that the statistics of the size of 
cascades remains the same. 

\begin{figure}[!h]
\begin{center}
\includegraphics[bbllx=30,bblly=20,bburx=380,bbury=330,scale=0.65]{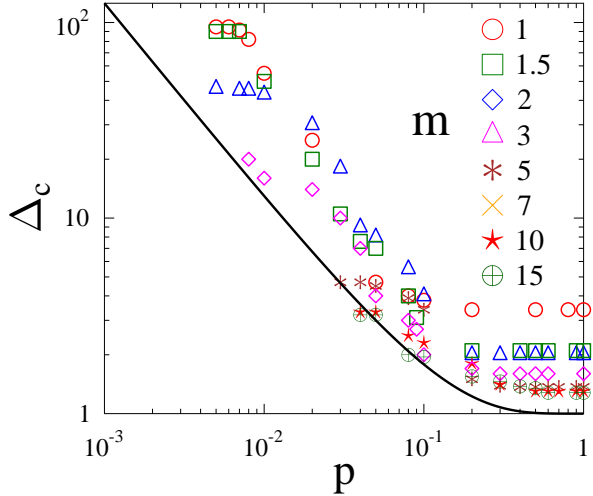}
  \caption{The symbols represent the crossover cascade size $\Delta_c$ obtained from the size distribution of cascades for several Weibull exponents $m$. 
  The analytic prediction (continuous curve) somewhat underestimates  $\Delta_c$, however,
  its functional form provides a reasonable description of the numerical findings.
   \label{fig:crosspoint}}
\end{center}
\end{figure}

It follows from the above arguments that the dependence of the LLS-ELS transition 
on the degree of strength disorder of the fibers (nodes) originates from the disorder dependence of the 
cascade activity of the system on the unperturbed regular lattice. 
For each Weibull exponent $m$ we estimated numerically the crossover point $\Delta_c$ of the cascade 
size distributions $p(\Delta)$ by determining the value of $\Delta$ where the two fitted straight 
lines of the two power law regimes of the distributions cross each other at each rewiring probability $p$. 
It can be observed in Fig.\ \ref{fig:crosspoint} that 
the analytical curve of $\Delta_c(p)$ obtained from Eq.\ (\ref{eq:crossover}) underestimates the numerical 
values, however, its functional form provides a reasonable description of the numerical findings.
In Figure \ref{fig:crosspoint} the rewiring probability 
of the lower $p_l$ and upper $p_u$ bounds of the transition at a given disorder $m$ 
can be identified as the $p$
values where the crossover first occurs, and where $\Delta_c$ becomes constant, respectively.
To obtain a more precise estimate of the transition regime, we solved numerically Eq.\ (\ref{eq:analitic_pl})
for $p_l$ substituting the value of the average largest cascade size $\left<\Delta_{max}\right>(p=0)$ at 
each $m$. This semi-analytical value of $p_l$ is compared in Fig.\ \ref{fig:disorder_all}$(d)$ to the
numerical one obtained as the position of the maximum of the relative scatter of $\Delta_{max}$ 
in Fig.\ \ref{fig:disorder_all} $(c)$.
The analytical results again underestimate the numerical ones but they have the same functional form. 
The value of $p_u$ we can estimate from the numerical results falls between 0.2 and 1.

\subsection{Failure triggered by low degree fibers}
Our simulations revealed that the competition of two mechanisms determine the behaviour of the loaded network 
both on the micro- and macro-scales. At higher rewiring probabilities $p$ the growing randomness of the network reduces the 
local load concentration in the system. This mechanism can substantially increase the cascade tolerance and the overall
load bearing capacity of the network especially at high disorder of the strength of nodes, see Fig.\ \ref{fig:disorder_all}$(a,b)$. However, as the degree 
distribution $\rho(k)$ broadens with increasing rewiring probability $p$, low degree nodes appear which have the counter effect 
of increasing the local load concentration when they fail. This effect becomes crucial at low node strength disorder, 
where the failure of a low degree node can easily trigger a catastrophic avalanche of failure events. 
In order to quantify this mechanism we characterize the degree of strength disorder of nodes by estimating 
the average of the smallest $\left<\sigma_{th}^{min}\right>$ and largest $\left<\sigma_{th}^{max}\right>$ failure thresholds in 
the bundle. Among $N$ independent random numbers sampled from the same probability distribution $P$,
the average of the smallest and largest values can be obtained as 
\beq{
\left<\sigma_{th}^{min}\right> = P^{-1}\left(\frac{1}{N+1}\right), \quad \mbox{and} \qquad 
\left<\sigma_{th}^{max}\right> = P^{-1}\left(1-\frac{1}{N+1}\right), 
}
where $P^{-1}$ denotes the inverse of the cumulative distribution \cite{hansen_statistical_2000}. 
Substituting the Weibull distribution 
Eq.\ (\ref{eq:expon_threshold}), the limit thresholds in a bundle of $N$ fibers can be cast into the form
\beq{
\left<\sigma_{th}^{min}\right> = \lambda\left(\frac{1}{N}\right)^{1/m}, \quad \mbox{and} \qquad 
\left<\sigma_{th}^{max}\right> = \lambda\left(\ln N\right)^{1/m}.
}
The ratio $r$ of the two values provides a measure of the degree of threshold disorder of the nodes 
\beq{
r=\left<\sigma_{th}^{max}\right>/\left<\sigma_{th}^{min}\right> = \left(N\ln N\right)^{1/m}.
\label{eq:ratio}
}

\begin{figure}
\begin{center}
\includegraphics[bbllx=60,bblly=30,bburx=380,bbury=330,scale=0.65]{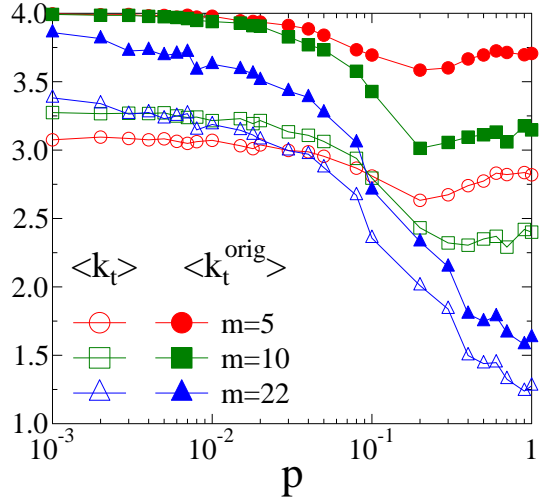}
  \caption{The average number of intact neighbors $\left<k_t\right>$ (empty symbols) and the average of 
  the original degree $\left<k_t^{orig}\right>$ (filled symbols) of the fiber the failure of which triggers 
  the final breakdown of the network of fibers. Results are presented for three values of the Weibull 
  exponent $m$. 
   \label{fig:trignode}}
\end{center}
\end{figure}

For the stability of the bundle the worst case is when a node of degree $k=1$ has the smallest 
failure threshold $\sigma_{th}^{min}$, since at failure it will double the load on its neighbor right 
at the beginning of the failure process. 
This load sharing will definitely result in failure of the neighbor if the elevated load $2\sigma_{th}^{min}$ is greater than the largest threshold $\sigma_{th}^{max}$ in the bundle so that the condition follows
\beq{
2\sigma_{th}^{min} > \sigma_{th}^{max}.
\label{eq:onedegree}
}
Since this secondary failure event gives rise to a large load increment on its
own neighbors, it is reasonable to assume that the cascade does not stop anymore and it becomes catastrophic. 
Equation (\ref{eq:onedegree}) implies that this mechanism determines the response of the bundle 
only when the strength distribution is sufficiently narrow $r<2$, and the rewiring probability $p$
is sufficiently high to have a finite fraction of nodes of degree $k=1$. Using the expression
of $r$ Eq.\ (\ref{eq:ratio}), the condition Eq.\ (\ref{eq:onedegree}) can be cast into a condition 
for the Weibull shape parameter 
\beq{
m>\frac{\ln \left(N\ln N\right)}{\ln 2},
\label{eq:condit_m}
}
which yields $m>20.9$ for the setup of our fiber bundle. 
It follows that the highest $m$ value we considered $m=22$ fulfills the condition 
so that the ultimate failure of the network at this $m$ should be dominantly triggered by fibers of degree 
1 at sufficiently high rewiring probabilities. To support the above arguments Fig.\ 
\ref{fig:trignode} presents the average number of intact neighbors $\left<k_t\right>$ of fibers 
the failure of which initiated the catastrophic cascade along with its original degree 
$\left<k_t^{orig}\right>$. 
It can be observed that both quantities monotonically decrease with increasing rewiring probability $p$, i.e.\
as the degree distribution of nodes gets broader with increasing $p$ the triggering node
has a lower and lower degree. At high strength disorder (low Weibull exponent $m$) the original 
$\left<k_t^{orig}\right>$ and final degrees $\left<k_t\right>$ of triggering fibers have a larger difference,
however, as the strength disorder gets reduced with increasing $m$ both curves shift to lower values
in such a way that their difference gets also smaller. It is important to emphasize that the 
$\left<k_t^{orig}\right>$ and $\left<k_t\right>$ curves tend to 1, which confirms that at low strength 
disorder the lowest degree nodes make the system vulnerable to cascading failure triggering the catastrophic breakdown of the system. The most remarkable outcome of these calculations is that the LLS-ELS transition 
is limited to a disorder range of the strength of nodes. At too low disorder rewiring makes the system more vulnerable to cascades which prevents any improvement of the strength and cascade tolerance of the system.
For Weibull exponents fulfilling the condition Eq.\ (\ref{eq:condit_m}) no LLS-ELS transition emerges.

\section{Discussion and conclusions}
We presented a theoretical study of the evolution of the failure dynamics of the fiber bundle model as the underlying network of load transmitting 
connections is gradually changed from a regular lattice to a random network. 
A complex network of fibers was constructed by randomizing a regular square lattice using the 
Watts-Strogatz rewiring technique. 
Fibers assigned to the nodes of the network are assumed 
to have a finite load bearing capacity which is a random variable. Initially all fibers are 
intact and their state is switched to failed when the local load on them exceeds their strength.
The system was subject to a slowly increasing external load by adding the same load increment to each
intact fiber in such a way that the failure of a single fiber is provoked. The load of failed fibers 
is transmitted to their intact nearest neighbors which may trigger an entire cascade of failure 
events under the constraint of load conservation. 

Gradually increasing the rewiring probability, we showed that the changing network structure 
gives rise to a transition from the localized to the mean field behaviour of failure processes accompanied by 
a complex evolution both on the macro- and micro-scales. 
The first deviations from the LLS behaviour of the regular lattice 
appear at a threshold probability $p_l$ where the transition sets on
and it gets completed by reaching the upper bound $p_u$, beyond which no further change occurs in the system. 
In the transition regime the probability distribution of the size
of failure cascades exhibits a crossover between two power laws of different exponents.
On the macroscopic scale the $\sigma(\varepsilon)$ 
curve of LLS bundles follow the mean field solution of the model at all rewiring probabilities 
but with a lower strength. The critical load and strain where ultimate failure occurs increase 
with the rewiring probability and tend towards limits which fall close to 
their mean field values. 

We demonstrated that the degree of disorder of the strength of fibers has a substantial effect 
on the transition: as the disorder gets reduced the 
transition regime shrinks and shifts to higher rewiring probabilities. Most notably the LLS-ELS 
transition is limited to a well-defined range of disorder of the strength of nodes. In particular, 
there exists a threshold amount of node strength disorder below which the randomization 
of the network of load transmitting connections does not provide any improvement neither 
of the overall load bearing capacity nor of the cascade tolerance of the system. 
Computer simulations revealed that at low strength disorder the fully random network is the most 
stable one, while at high disorder best cascade tolerance is obtained at a lower structural randomness.

Based on the interplay of the network structure and node strength disorder we constructed an analytical 
argument which provided a reasonable description of the numerical findings. These calculations revealed 
that two competing mechanisms determine the response of the network: the rewiring of the underlying lattice
introduces long range random connections in the load transmission network which reduce the load concentration 
around failed regions, and in turn allow the system to tolerate larger cascades without becoming instable.
At low rewiring probabilities due to the low fraction of long range contacts, small sized cascades
remain practically unaffected. However, beyond a characteristic size, cascades involve more and 
more fibers with rewired contacts which increase their stability. This mechanism leads to 
the increase of the cutoff cascade size and the emergence of a second power law regime of the distributions
with a lower exponent than for the small cascades. 
However, as the rewiring probability increases the degree distribution broadens which increases the fraction 
of low degree nodes. This gives rise to the counter effect that in the vicinity of failing low degree 
nodes a large load concentration emerges, which can trigger catastrophic cascades. As a consequence, at high 
rewiring probabilities the network becomes vulnerable to early cascades when the strength of nodes is sampled
from a narrow distribution.

The failure mechanism of the fiber bundle model we focused on is quite generic with four key elements:
$(I)$ the total load on the system
is increased by adding the same load increments to all the intact elements;
$(II)$ nodes fail irreversibly such that they are removed from the bundle together with their links; 
$(III)$ failed nodes transfer their load to their intact nearest neighbors through their links; 
$(IV)$ the load on the system is conserved during the spreading of failure cascades. The fiber bundle model 
has been used to study the emergence of cascading breakdown of roads carrying traffic, flow channels, 
and power grids. Due to the minimum amount of additional assumptions we made, our results should be 
relevant for these modelling approaches as well.

\section*{Acknowledgments}
The work is supported by the EFOP-3.6.1-16-2016-00022 project. 
The project is co-financed by the European Union and the European Social Fund.
This research was supported by the National Research, Development and
Innovation Fund of Hungary, financed under the K-16 funding scheme 
Project no.\ K 119967.
The research was supported by the Thematic Excellence Programme (TKP2020-NKA-04) of the Ministry 
for Innovation and Technology in Hungary.

\bibliographystyle{elsarticle-num} 

\bibliography{/home/feri/papers/statphys_fracture}

\end{document}